\documentclass[floatfix,aps,showpacs]{revtex4}
\usepackage{graphicx}

\begin{document}

\newcommand{\x}{{\bf x}}

\title{Thermodynamics of a trapped interacting Bose
gas and the renormalization group}
\author{G.~Metikas, O.~Zobay, and G.~Alber}
\affiliation{Institut f\"ur Angewandte Physik, Technische Universit\"at 
Darmstadt, 64289 Darmstadt, Germany}
\begin{abstract}

We apply perturbative renormalization group theory to the symmetric phase of a dilute interacting Bose gas which is trapped in a three-dimensional harmonic potential. Using Wilsonian energy-shell renormalization and the $\varepsilon$-expansion, we derive the flow equations for the system. We relate these equations to the flow for the homogeneous Bose gas. In the thermodynamic limit, we apply our results to study the transition temperature as a function of the scattering length. Our results compare well to previous studies of the problem.

\end{abstract}

\maketitle

\flushbottom

\section{Introduction}

Dilute, repulsively interacting Bose gases in a three-dimensional harmonic
 trap have been at the focus of attention of the theoretical physics community
since the experimental realization of Bose-Einstein condensation (BEC)
in 1995 \cite{AndEnsMat95,DavMewAnd95,BraSacTol95}. A vast amount of literature has been published recently on this topic
based on the mean-field approach. It is however known that mean-field
 theory is not appropriate near the critical temperature because the
fluctuations dominate the mean field in this region. This follows from
applying the so-called Ginzburg criterion \cite{Kadanoff, Huang}
and has triggered
several attempts to go beyond mean-field theory near the critical region
by means of various renormalization group (RG) techniques
in the homogeneous interacting
Bose gas \cite{StoofBijlsmaRen, BraatenNieto, Andersen,
Alber, Crisan, AlberMetikas, Metikas}. This recent work was added to
a significant amount of existing literature on the application of RG methods to interacting
Bose gases, that was written when such systems were only an
interesting theoretical problem \cite{CreswickWiegel, FisherHohenberg,
Weichman, Nelson, Kolomeisky1, Kolomeisky2}.

All these studies so far concern the homogeneous case; the real
challenge however is to apply renormalization techniques to the experimentally
relevant system, the trapped Bose gas \cite{Andersen}. This challenge is met
here for the first time, at least when we approach the transition region from
 above the critical temperature (symmetric phase). In this
case the theoretical treatment is simpler than in the symmetry-broken
phase and allows us to see how to apply renormalization techniques. In
particular, we shall use a method similar to what in the homogeneous case is
known as momentum-shell renormalization \cite{StoofBijlsmaRen, Andersen,Wilson}. In
the trapped case it is energy shells instead of momentum shells that we will be
integrating out. Of course, since we are dealing here with a finite system,
there is, strictly speaking, no phase transition
\cite{Kadanoff}. However, as we approach the thermodynamic
limit, a quasi-phase transition develops, and
 renormalization methods are expected to enable us to study non-universal
properties
such as the critical temperature and its dependence on
the scattering length.

The paper is organized as follows.
Section II develops the theoretical methods necessary to apply the
renormalization procedure to the weakly interacting, trapped Bose gas. Starting from a high-energy
 cutoff, we successively integrate out energy shells thus perturbatively
creating an
 effective action for the low-energy field. This effective action is then cast
 into the form of the original action.
In Sec.\ III, we derive the renormalization
group equations for the chemical potential, the interaction, and the
grand-canonical thermodynamic potential.
We also refine
our results by means of a first-order $\varepsilon$-expansion. It is shown that, in the thermodynamic limit, the flow equations for the chemical interaction and the interaction have the same form as for a homogeneous interacting Bose gas. All nonuniversal properties, that are due to the presence of the harmonic trap, enter through the flow equation for the thermodynamic potential.
Section IV is devoted to the study of a particular
non-universal property, namely the transition temperature and its dependence on
the scattering length in the thermodynamic limit. After a short summary in Sec.\ V,
we discuss in Appendix A how  the thermodynamic limit affects
the relevant density of states entering the renormalization group equations of
Sec.\ III.
Appendix B contains a discussion of numerical methods we used for obtaining the
results presented in Sec.\ IV.

\section{Effective action for the low-energy field}

The grand partition function for the interacting dilute Bose gas can be
expressed as a functional integral  \cite{Feynman,NotesStoof}, i.e.,
\begin{equation}
Z = \int \delta[ \phi, \phi^{*}]~ e^{- \frac{1}{\hbar}  S[\phi,\phi^{*}]}.
\label{partfunc}
\end{equation}
We work in a $D$-dimensional space, where
the bosonic fields $\phi(\tau, \x)$
and $\phi^{*}(\tau, \x)$ depend on
spatial coordinates $\x=(x_{1}, ... , x_{D})$ and on imaginary time
$\tau$. The fields are periodic in $\tau$ with period $\hbar \beta$, where
$\beta = 1/k_{B} T$ is the inverse temperature and $k_{B}$ denotes Boltzmann's constant.
Assuming $s$-wave repulsive interactions, the Euclidean action characterizing the functional integral
of Eq.\ (\ref{partfunc}) is
\begin{eqnarray}
S[\phi, \phi^*] &=&  \frac{1}{\hbar}
 \int_0^{\hbar \beta }\! d\tau\!\int \! d^D {\bf x}
\left\{ \phi^* (\tau, {\bf x})
\left[\hbar\frac{\partial}{\partial \tau} - \frac{\hbar^2}{2m} \nabla^{2} +
 V({\bf x})      - \mu\right]
 \phi(\tau, {\bf x}) + \frac{g}{2} |\phi(\tau, {\bf x})|^4
 \right\}{\ }
\label{action}
 \end{eqnarray}
 where $\mu$ is the chemical potential, $m$ the particle mass, and the spatial coordinates are integrated over all space.
The coupling constant for the short-range, repulsive interaction potential
is denoted by $g$.
 For a dilute Bose gas, it has to renormalize to the two-body
 $T$-matrix when all two-body scattering processes are taken into account
\cite{FetterWalecka, NotesStoof, StoofBijlsmaRen}.
Thus,
in three spatial dimensions, at zero energy and in lowest order perturbation theory, this
coupling constant is related to the $s$-wave scattering length $a$ by
$g = 4\pi a \hbar^2/m$. An improved relation for the scattering length $a$ is discussed later
[compare with Eq.\ (\ref{scatteringlength})].

We assume that the interacting Bose gas is trapped in a
$D$-dimensional isotropic harmonic oscillator potential
with frequency $\omega$, i.e., $V({\bf x}) =\frac{1}{2} m \omega^2 {\bf x}^2.$
Furthermore let us define the trapping potential along each of its spatial
dimensions as $V_j(x_{j})= m \omega^2 x_{j}^2/2, j=1,\dots,D$. The
corresponding normalized eigenfunctions are denoted by
$\psi_{ n_{j}}(x_{j})$, $n_j\ge 0$, and have eigenenergies
$E_{n_{j}} + E_{0}/D$, where
$E_{n_{j}}=\hbar \omega n_{j}$ and $E_{0}=D \hbar \omega /2$.
The $D$-dimensional eigenfunctions are given by
 $\psi_{n_{1},\dots,n_{D}}(\x ) =
\psi_{n_{1}}(x_{1}) \dots \psi_{n_{D}}(x_{D})$
with eigenenergies $E_{n}+E_{0}=\hbar \omega n + E_0$ where
$n = n_{1} + \dots + n_{D} $.

We now wish to apply a modification of the momentum-shell RG
procedure to the trapped interacting Bose gas characterized
by the effective action of Eq.\ (\ref{action}).
In the homogeneous case the momentum-shell method has been explored extensively
\cite{StoofBijlsmaRen, BraatenNieto, Andersen,
Alber, Crisan, AlberMetikas, Metikas}. In the case we are dealing with here, the
noninteracting Hamiltonian ($g=0$) derived from (\ref{action}) does not
commute with the momentum operator because of the presence of the
trapping potential. We will therefore use harmonic oscillator
eigenstates instead of momentum eigenstates and
integrate out small energy shells.

For the purpose of evaluating the partition function (\ref{partfunc}),
we impose a high-energy cutoff by assuming that the maximum value that $n$ can
take is some large integer $ n_{\Lambda}\gg 1$. We define an energy shell as the
shell between $  n_{\Lambda} - \delta n_{\Lambda}$ and $n_{\Lambda}$,
where $\delta n_{\Lambda}$ is an integer such that $\delta
n_{\Lambda} / n_{\Lambda} \ll 1$.
In addition, we split the bosonic field
into low-energy and high-energy components denoted by $\phi_{<}$
and $\phi_{>}$, respectively, so that $\phi(\tau, \x ) = \phi_{<}(\tau,\x ) +
\phi_{>}(\tau, \x ).$ Thus, the total bosonic field expanded on eigenfunctions of the noninteracting Hamiltonian is
\[ \phi(\tau, \x )=
\sum_{l=-\infty}^{\infty} \sum_{n=0}^{n_{\Lambda}}
\sum^{'}_{n_{1} \dots n_{D}}
\frac{ e^{i p_{0}^{l} \tau } }{ \sqrt{\hbar \beta}}
~\psi_{n_{1},\dots,n_{D}}(\x )~
{\tilde \phi}(l,n_{1},\dots,n_{D})\]
where $p_{0}^{l}= 2 \pi l / \hbar \beta$ are the Matsubara frequencies,
${\tilde \phi}(l,n_{1},\dots,n_{D})$ are the complex-valued expansion coefficients,
 and the prime on the sum over $n_1, \dots, n_D$ signifies the constraint
$n_{1} \geq 0,\dots ,n_{D} \geq 0,~ n_{1} + \dots + n_{D}=n.$
Correspondingly,
the low-energy field is
\[ \phi_{<}(\tau, \x )=
\sum_{l=-\infty}^{\infty} \sum_{n=0}^{n_{\Lambda} - \delta
n_{\Lambda}}
\sum^{'}_{ n_{1} \dots n_{D} }
\frac{ e^{i p_{0}^{l} \tau } }{ \sqrt{ \hbar \beta}}~
\psi_{n_{1}, \dots ,n_{D}}(\x )~
{\tilde \phi}(l,n_{1},\dots,n_{D}),\]
and the high-energy field is
\[ \phi_{>}(\tau, \x )=
\sum_{l=-\infty}^{\infty} \sum_{n=n_{\Lambda}-\delta n_{\Lambda}
}^{n_{\Lambda}}
\sum^{'}_{n_{1} ... n_{D}}
\frac{ e^{i p_{0}^{l} \tau } }{ \sqrt{\hbar \beta} }~
\psi_{n_{1},...,n_{D}}(\x )~{\tilde \phi}(l,n_{1},...,n_{D}). \]

The first stage of the RG procedure is usually referred to as Kadanoff
transformation and consists of two steps; first, an effective action for the
low-energy field is derived, and subsequently this effective action is cast
in the form of the original action.
For this purpose
we proceed to the one-loop calculation of the effective theory for
the low-energy field by integrating out the high-energy field. Analogously to
the homogeneous case \cite{CreswickWiegel}, we thus obtain an effective action of the form
\begin{equation}
S_{\rm eff}[\phi_{<}, \phi_{<}^{*}] = S[\phi_{<}, \phi_{<}^{*}] +
{\textstyle \frac{1}{2}} {\rm Tr}[\ln(1 - \hat{G}^{>} \hat{\Sigma})].
\label{kadaction}
\end{equation}
The ``Tr" symbol  denotes the trace in 
both the functional and the internal space of 
$ \hat{G}^{>} \hat{\Sigma} $. 
These latter quantities denote the
Green's function for the high-energy field, i.e.,
 \[ \hat{G}^{>}(\hat{p}_{0}, \hat{H}_{{\rm ho}})  =
\left( \begin{array}{cc}
                  \hat{B}(\hat{p}_{0}, \hat{H}_{{\rm ho}}) & 0 \\
                  0    &   \hat{B}^{*}(\hat{p}_{0}, \hat{H}_{{\rm ho}})
                  \end{array} \right), \] \\
and the self-energy for the low-energy field, i.e.,
 \[ \hat{\Sigma}(\hat{\tau}, \hat{\x }) = g \left( \begin{array}{cc}
                   2 \phi^{*}_{<}(\hat{\tau}, \hat{\x })
\phi_{<}(\hat{\tau}, \hat{\x })  & 
\phi_{<}(\hat{\tau}, \hat{\x }) \phi_{<}(\hat{\tau}, \hat{\x }) \\
  \phi_{<}^{*}(\hat{\tau}, \hat{\x}) \phi_{<}^{*}(\hat{\tau}, \hat{\x }) &
 2 \phi^{*}_{<}(\hat{\tau}, \hat{\x }) \phi_{<}(\hat{\tau}, \hat{\x })
                  \end{array} \right), \] \\
with
\begin{equation}
 \hat{B}( \hat{p}_{0}, \hat{H}_{{\rm ho}} ) =
 \frac{1}{i \hat{p}_{0} + \hat{H}_{{\rm ho}}},~~
 \hat{H}_{{\rm ho}} =
\frac{\hat{{\bf p}}^2}{2m} + V(\hat{\x })- \mu.
\end{equation}
The hat indicates that these quantities are Schwinger-Fock operators
\cite{SchwingerFock,Aitchison,AML}.

As in the homogeneous case, we now perform an
expansion of (\ref{kadaction}) up to second order in
$ \hat{G}^{>} \hat{\Sigma} $.
This expansion is particularly well justified in
 the context of the RG procedure used here, because the
 self-energy is multiplied by the high-energy propagator which is inversely
 proportional to the largest energy scale of the system, the cutoff energy
 $\hbar \omega n_{\Lambda}$. Thus an RG perturbative expansion is expected
 to have a wider range of validity than ordinary perturbation as employed, for
 example, in the mean-field theory context.
 Furthermore, the truncation at
 second order, that is at quartic interactions, is
self-consistent with the truncation at quartic interactions of the
original action (\ref{action}). Higher order terms in powers of the
low-energy field are discarded
 exactly as in the
homogeneous case. A further discussion of these terms that we neglect here can
be found in \cite{Metikas, Andersen}. The second-order expansion yields
\begin{equation}
 {\rm Tr} [\ln (1-\hat{G}^{>} \hat{\Sigma})] \approx {\rm Tr}
[ -\hat{G}^{>} \hat{\Sigma} - {\textstyle \frac{1}{2}} (\hat{G}^{>} \hat{\Sigma})^2].
\label{trace}
\end{equation}
Performing the sums over the Matsubara frequencies we find for the
 first trace
\begin{equation}
{ \rm Tr }[ \hat{G}^{>} \hat{\Sigma}]= \int_{0}^{\hbar \beta} d\tau
\int d^{D} \x ~|\phi_{<}(\tau,\x )|^2~2g
\sum_{n=n_{\Lambda}- \delta n_{\Lambda} }^{n_{\Lambda}}
\sum^{'}_{n_{1} ... n_{D}} f_1(E_n+E_0)|\psi_{n_{1} ... n_{D}}(\x )|^2
\label{firsttrace}
\end{equation}
 and for the second trace
\begin{equation}
{\rm Tr} [ \hat{G}^{>} \hat{\Sigma} \hat{G}^{>} \hat{\Sigma}] =
\int_{0}^{\hbar \beta} d\tau \int d^{D} \x ~|\phi_{<}(\tau, \x)|^4~ 2 g^2
\sum_{n = n_{\Lambda}- \delta n_{\Lambda} }^{n_{\Lambda}}
\sum^{'}_{n_{1} ... n_{D}} f_2(E_n+E_0)
|\psi_{n_{1} ... n_{D}}(\x )|^2,
\label{secondtrace}
\end{equation}
 where
 \begin{eqnarray}
f_{1}(E)&=& 1+ 2 N_{BE}(E), \nonumber \\
f_{2}(E)&=& 4 \beta N_{BE}(E) [1+N_{BE}(E)] +
\frac{1+2N_{BE}(E)}{2(E-\mu)},
\label{f}
\end{eqnarray}
and
$N_{BE}(E)= [e^{\beta (E - \mu)} -1]^{-1}$
is the Bose-Einstein distribution.

We proceed to the second step of the Kadanoff transformation noting
that the first trace, Eq.\ (\ref{firsttrace}), is quadratic and the second trace,
 Eq.\ (\ref{secondtrace}), is quartic
in the modulus of the low-energy field.  
Therefore the first trace can be interpreted as 
a correction to the quadratic part of the
original action (\ref{action}), 
\begin{equation}
d(\mu - V) = - g \sum_{n=n_{\Lambda}- \delta n_{\Lambda}
}^{n_{\Lambda}} \sum^{'}_{n_{1} ... n_{D}} f_1(E_n+E_0)
|\psi_{n_{1} ... n_{D}}(\x )|^2,
\label{mu}
\end{equation}
and the second trace as a correction to the quartic part of (\ref{action}),
\begin{equation}
dg=-g^2 \sum_{n=n_{\Lambda}- \delta n_{\Lambda} }^{n_{\Lambda}}
\sum^{'}_{n_{1} ... n_{D} } f_2(E_n+E_0)
|\psi_{n_{1} ... n_{D}}(\x )|^2.
\label{g}
\end{equation}

\begin{figure}
\includegraphics[width=8.6cm]{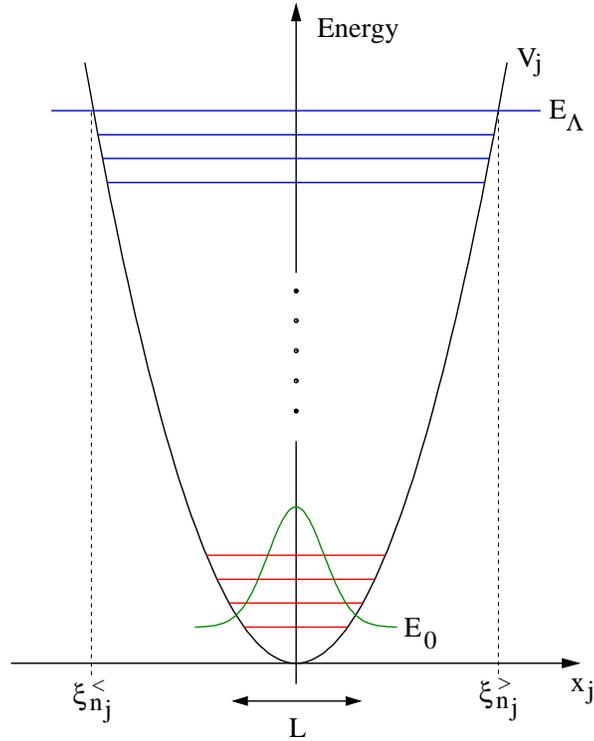}
\caption[f2]{
Schematic for integrating out a high-energy shell. Shown is a one-dimensional cut through the potential along the coordinate axis $x_j$. The high-energy eigenfunctions are located around $E_{\Lambda}\gg E_0$.
In Eq.\ (\protect\ref{wkb}), the trap potential $V_j$ is thus negligible compared
to $E_{\Lambda}$ around the trap center.}
\label{fig1}
\end{figure}

We observe that the corrections (\ref{mu}) and (\ref{g})
are $\x$-dependent and therefore
not of the form of the original action (\ref{action}). However, in our subsequent
treatment we are mainly interested in
the limit of small trapping frequencies, i.e., $\beta\hbar \omega \ll 1$.
In this case, the thermal wave length $\lambda_{th} = \sqrt{2\pi \hbar^2\beta/m}$ is
small in comparison with the characteristic extension $L = \sqrt{\hbar/m\omega}$ of the ground state of the harmonic
trap. Therefore, it is expected that the thermodynamic behavior
is dominated by the properties of the interacting Bose gas in the center of the trap and boundary effects
are negligible.
In addition, this limit is consistent with the thermodynamic limit which we will concentrate on later and where $\omega$ and the particle number $\tilde{N}$ tend to $0$ and $\infty$, respectively,
such that $\tilde{N}\omega^3$ remains constant.
Thus, let us focus on the region near the center of the trap
which means $V_j(x_{j} ) \ll E_{n_j} + E_{0}/D, \ j=1,...,D$.
Figure \ref{fig1} shows a schematic of the situation under consideration. We now employ the JWKB approximation
for the harmonic oscillator eigenfunctions
\cite{BerryMount,Gleisberg}, i.e.,
\begin{equation}
\label{wkb}
\psi_{n_{j}}(x_{j}) =
 \frac{\sqrt{\frac{ \omega}{\pi}}
(2m)^{1/4} }{ \left[ E_{n_{j}} + E_{0}/D - V_j(x_{j}) \right] ^{1/4}}
\cos \left[  \frac{1}{\hbar} \int_{\xi^<_{n_j}}^{x_j}
dy \left(  \sqrt{2m[ E_{n_{j}} + E_{0}/D - V_j(y)]} -
\frac{\pi}{4} \right) \right],
\nonumber
\end{equation}
with $\xi^{>}_{n_{j}} = L  \sqrt{2n_{j}+1}$ and  $\xi^{<}_{n_{j}} = -L  \sqrt{2n_{j}+1}$ denoting the right and left classical turning points. Expression (\ref{wkb}) is valid in the classically allowed region $\xi^{<}_{n_{j}}\le x_j  \le \xi^{>}_{n_{j}}$ except for a small area near the turning points.
In particular, near the trap center it is possible to neglect the trapping potential $V_j(x_{j})$ in the denominator of (\ref{wkb}).
Furthermore, in this region the high-energy eigenstates oscillate very rapidly on the length scale set by $L$.
Therefore, we can safely approximate each of the squares of the cosines
appearing in (\ref{mu}) and (\ref{g}) by their spatial averages $1/2$. The corrections in Eqs. (\ref{firsttrace}) and (\ref{secondtrace}) now assume the forms
\begin{eqnarray}
d(\mu - V) &=&
 -g  \sum_{n=n_{\Lambda}- \delta n_{\Lambda}
}^{n_{\Lambda}} \rho(n) f_{1}(E_{n}+E_{0})
\frac{(2 m \omega)^{D/2}}{\hbar^{D/2} (2 \pi)^D}
\label{wkbmu}
\end{eqnarray}
and
\begin{eqnarray}
 dg &=&
-g^2 \sum_{n=n_{\Lambda}- \delta n_{\Lambda}
}^{n_{\Lambda}} \rho(n) f_{2}(E_{n} + E_{0})
  \frac{(2 m  \omega)^{D/2}}{\hbar^{D/2} (2 \pi)^D},
\label{wkbg}
\end{eqnarray}
where 
\begin{equation}
\rho(n)=
\sum^{'}_{n_{1} ... n_{D}} \frac{1}{ \sqrt{(n_{1}+\frac{1}{2})
 ... (n_{D}+\frac{1}{2}) } }.
\label{rho}
\end{equation}

We note that because we are focusing on the center of the trap, there
is no explicit dependence on the trapping potential
 anymore on the right-hand side of (\ref{wkbmu}) and (\ref{wkbg}).
We can therefore interpret the right-hand side of
(\ref{wkbmu}) as a correction to the chemical potential only and set
$dV=0$. In other words, the trapping potential and consequently the
trapping frequency receive no non-trivial corrections and exhibit
trivial scaling only.

\section{Renormalization group equations}
\subsection{Derivation of the flow equations}
At this stage we replace the sum over $n$ in Eqs.\ (\ref{wkbmu}) and (\ref{wkbg}) by
an integral over $n$. It is well known that, as long as the functions we are
summing over are smooth, this
is a satisfactory approximation \cite{BerryMount, Grossmann1, Grossmann2}, and it is
particularly good in the regime we are considering here,
$ \beta \hbar \omega \ll 1$
\cite{KetterlevanDruten}.
We can now consider the renormalization step to be infinitesimal
and pursue the analogy with the homogeneous calculation.
Therefore, we shift the origin of integration over $n$ in (\ref{wkbmu})and
(\ref{wkbg}), i.e. $n \to n -\frac{D}{2}$, thus eliminating the zero-point
energy from the functions $f_{1}, f_{2}$ in (\ref{f}).

In order to emphasize the analogy to the case of a homogeneous interacting
Bose gas
we introduce a generalized momentum vector
${\bf p}=(p_{1},...,p_{D})$  through the relation
\begin{equation}
\hbar \omega n_{j} = \frac{p_{j}^2}{2m},~~ p_{j}
\geq 0, ~~ j=1,...,D.
\label{cgenp}
\end{equation}
Obviously,
the modulus $p$ of this generalized momentum is given by
$\hbar \omega n = p^2/2m$.
We can also define a cutoff $\hbar \Lambda$ for the generalized momentum vector by
the relation
$\hbar \omega n_{\Lambda}  = (\hbar \Lambda)^2 /2m = E_{\Lambda}$.
Correspondingly, $1/\Lambda$ may be viewed as the smallest
length in our problem, i.e.,
$\Lambda^{-1} \ll \lambda_{th}$ or equivalently $\beta_{\Lambda}  \ll \beta$
with $\beta_{\Lambda}=m/\hbar^2 \Lambda^2=1/2 E_{\Lambda}$.
As in the homogeneous case, we may parametrize the generalized momentum in terms of
this cutoff and of a
dimensionless continuous parameter $l$ according to
$p/\hbar =\Lambda e^{-l}$.
Thus, differentiating Eqs.\ (\ref{wkbmu}) and (\ref{wkbg})
with respect to $l$ we obtain the flow equations
\begin{eqnarray}
\frac{d\mu }{dl} & = &
-g~(\Lambda~e^{-l})^D~d(l)~
f_{1}(E_{\Lambda} e^{-2l}), \nonumber \\
\frac{dg}{dl} &=&
-g^2~(\Lambda~e^{-l})^D~
d(l)f_{2}(E_{\Lambda} e^{-2l})
\label{flows}
\end{eqnarray}
with
\begin{equation}
d(l) = \frac{2^{D/2}}
{(2 \pi)^D}(L\Lambda~e^{-l})^{2-D}\rho[(L\Lambda~e^{-l})^2/2 - D/2].
\label{density}
\end{equation}
This latter factor describes all effects originating in the discrete energy level structure in the isotropic
harmonic trap. In the case of a homogeneous interacting Bose gas it assumes the constant
value $d_h = 1/2\pi^2$
for the physically relevant case of $D=3$.

The effective action of Eq.\ (\ref{kadaction}) can now be cast in the form of the original action.
This is achieved by reinstating the
cutoff for $n$ to its original value $n_{\Lambda}$, or, equivalently, the cutoff
of the modulus of the generalized momentum $q$ of the low-energy field to its
original value $\hbar\Lambda$ (note that we are using $q$ for the modulus of the
momentum of the low-energy
 field as opposed to $p$ which is the modulus of the momentum of the
high-energy field). To this end,
 we rescale each component $q_{j}$ of the momentum vector ${\bf q}=
 (q_{1},...,q_{D})$ of the low-energy field according to
$q_{j}(l) = q_{j} e^{l}$. Thus, the rescaled modulus of the momentum vector of
the low-energy field reads $q(l) = q e^{l}$ and it is cut off at $\hbar\Lambda$.
This trivial rescaling together with
 the demand that the effective action (\ref{kadaction}) remains formally the
 same after each renormalization step induces the trivial scaling of the
 parameters of the effective action, $x(l)=x e^{-l}$,
$\tau(l)= \tau e^{-2l}$,  $ \mu(l)=\mu e^{2l}, g(l)=g e^{(2-D)l},
\beta(l) = \beta e^{-2l}, \phi(l)= \phi e^{D l /2}$.
We observe that all quantities scale as in the
homogeneous case. In addition, we define here the trivial scaling of the
oscillator frequency, $\omega(l) = \omega e^{2l}$.
Consequently, the trivial scaling of
the oscillator length is given by
$L(l)=L e^{-l}$.
It is convenient to introduce the dimensionless parameters $b(l) = \beta(l)/
\beta_{\Lambda}, M(l) =\beta_{\Lambda} \mu(l),
{\tilde G }(l) = \beta_{\Lambda} \Lambda^D g(l)/b(l), \Omega(l) = \hbar\beta_{\Lambda}\omega(l)=(L\Lambda~e^{-l})^{-2}$ and the dimensionless cutoff energy
$E_{>} \equiv\beta_{\Lambda}E_{\Lambda} = 1/2$.
We want to point out that the dimensionless quantities $b(l)$ and $M(l)$ 
scale trivially
as the corresponding dimensional parameters $\beta(l)$ and $\mu(l)$,
whereas ${\tilde G}(l)$ scales trivially as
${\tilde G}(l)={\tilde G} e^{(D-4)l}$.
In terms of these quantities the renormalization group equations for the dimensionless chemical potential
and the dimensionless coupling strength are finally given by
\begin{eqnarray}
\frac{dM(l)}{dl}&=&2 M(l)- {\tilde G}(l)d(l)b(l)
\left[ 1+ 2 N(l) \right], \nonumber \\
\frac{d{\tilde G}(l)}{dl}&=& (4-D){\tilde G}(l)- {\tilde G}(l)^2~d(l)b(l)
\left\{ 4~b(l)~N(l)~[1+N(l)] + \frac{1+ 2 N(l)}{2 (E_{>} - M(l))} \right\}
\label{trgeq}
\end{eqnarray}
with the scaled Bose distribution
\begin{equation}
N(l)=\left\{ e^{b(l) [E_{>}-M(l)]}-1 \right\}^{-1}.
\label{scaledBose}
\end{equation}
We observe that there are two differences from the homogeneous RG
equations.

The first difference is that the maximum number of RG steps
in the trapped case is finite. This is due to the presence of the
zero-point energy which has no homogeneous counterpart.
The number of renormalization steps
$l$ required to reach an energy scale $E(p)\equiv p^2/2m=\hbar\omega n$ is defined by
$E(p) e^{2l}= E_{\Lambda}$ and therefore $l=\ln
\sqrt{E_{\Lambda}/E(p)}$. This means that, if $E_{{\rm min}}$ is the minimum
value the energy can take, the maximum value of RG steps is $l^{*}=
\sqrt{E_{\Lambda}/E_{{\rm min}}}$. In the homogeneous case $E_{{\rm
min}}=0$
 and $l^{*}=\infty$ whereas in the trapped case $E_{{\rm min}}=
D \hbar \omega/2$, which results in a finite number of steps,
\begin{equation}
l^{*}=\ln{ \frac{1}{\sqrt{D \beta_{\Lambda}\hbar\omega}}} = \ln{ \frac{1}{\sqrt{D \Omega(0)}}}.
\label{l*}
\end{equation}
Of course,
this maximum number of steps
must be consistent with the requirement that throughout the renormalization
procedure the Bose-Einstein distribution of Eq.\ (\ref{scaledBose})
must remain positive so that
the particle density remains positive.

The second difference is due to the fact that in the trapped case the
energy spectrum is discrete. Although we replaced the sum over $n$ by
an integral over $n$ we kept the sums in the expression for
$\rho(n)$ of Eq.\ (\ref{rho}) and thus retained part of the discrete nature of the problem.
This method is also employed in treatments of the non-interacting trapped Bose gas
\cite{Grossmann1, Grossmann2} where it is also shown
that, for $D=3$,
the corrections due to the discreteness play a significant role
only for small particle numbers, typically $\tilde{N} \leq 1000$.
The consequence of retaining the sums in $\rho(n)$ is that the quantity $d(l)$ of Eq.\ (\ref{density})
appearing in the RG equations (\ref{trgeq}) differs from the
homogeneous case in two respects. First, it depends on the
renormalization step $l$ whereas the corresponding homogeneous
quantity $d_{{\rm h}}$ is a constant, e.g., for $D=3$ we have $d_{{\rm h}}=1/2 \pi^2$.
 Furthermore, $d(l)$ does not coincide with the density of trapped
states, whereas $d_{{\rm h}}$ does coincide with the density of
homogeneous states \cite{StoofBijlsmaRen, Metikas}.

\subsection{Thermodynamic limit}

In our subsequent discussion we are particularly interested in the thermodynamic limit
of the RG flow equations. As mentioned above, in the case of an isotropic harmonic trap this limit is defined
by letting the trapping frequency $\omega$ tend to zero and the number of particles $\tilde{N}$
to infinity in such a way that $\tilde{N}\omega^3$ remains finite.
It is shown in Appendix A that
in this limit the quantity $d(l)$ approaches the constant value $1/2\pi^2$ in three
spatial dimensions. We thus conclude that in the thermodynamic limit the
RG flow equations (\ref{trgeq}) for the trapped interacting gas assume the same form as in the homogeneous case. This is one of the main results of this work. In particular, it follows that the universal critical properties are not influenced by the presence of the trap. As we will see below,
all effects on the (nonuniversal) thermodynamic properties that originate from the isotropic harmonic trap are contained
in the flow equation for the grand thermodynamic potential of Eq.\ (\ref{w}) below. It involves
the density of states in the trap and therefore differs from the
corresponding flow equation for a homogeneous Bose gas.

\begin{figure}
\begin{center}
\includegraphics[width=6.8cm]{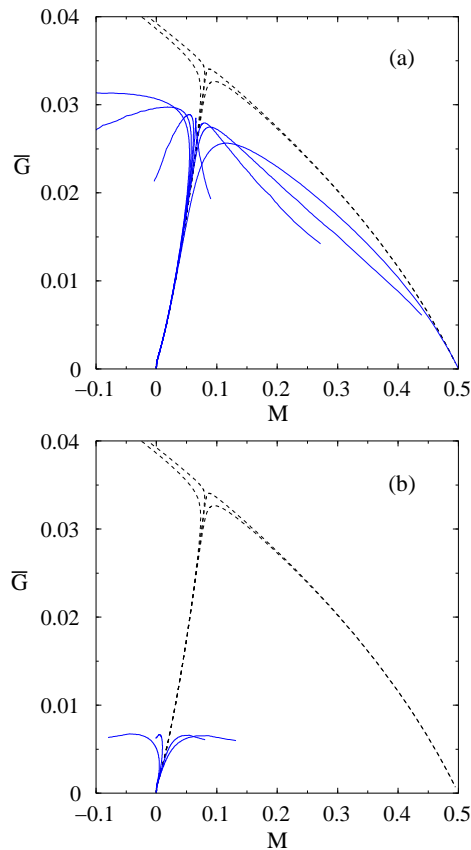}
\caption{ Flow trajectories determined from the trapped RG
equations (\ref{trgeq}) (full curves) and, in the thermodynamic limit
($\omega \to 0$, ${\tilde N} \to \infty$, ${\tilde N} \omega^3 = \
{ \rm constant} $),
the corresponding
homogeneous trajectories (dashed curves). All trajectories have the same
initial values of
$b(0)=10657$ and $G(0)=4\pi$ but different values of
$M(0)$. For the homogeneous trajectories  $M(0)$ varies
between $2.1387 \times 10^{-6}$ and $2.1391 \times 10^{-6}$.
The corresponding value for $a \tilde N^{1/6}/L$ is $6.12 \times
 10^{-3}$. The trapped trajectories in (a) are for $\Omega(0) \equiv 1/(L\Lambda)^2 = 10^{-
9}$
and $M(0)$ between $2.1195 \times 10^{-6}$ and $2.128 \times 10^{-6}$.
The trapped trajectories in (b) are for $\Omega(0) = 10^{-6}$ and $M(0)$
between $1.4 \times 10^{-6}$ and $2.1 \times 10^{-6}$. The quantity $\bar G$ is related to $\tilde G$ by ${\bar G}={\tilde G}d(l)$.
\label{fig2}}
\end{center}
\end{figure}

It is also instructive to study the transition to the thermodynamic limit in terms of the flow resulting from the RG equations (\ref{trgeq}). Figs.\ \ref{fig2}(a) and (b) compare typical flow trajectories close to and further away from the thermodynamic limit. The proximity to this limit is determined by the value of $\Omega(0)=1/(L\Lambda)^2$ which differs by several orders of magnitude between the two plots. Within each of the diagrams, however, the trajectories were obtained for fixed initial values of $b$, ${\tilde G}$, and $\Omega$, but differing values of the scaled chemical potential $M(0)$. The dashed curves show representative trajectories for the corresponding homogeneous flow, i.e, $\Omega(0)=0$. These trajectories clearly reflect the presence of the unstable fixed point in the homogeneous flow which is indicative of a second-order phase transition. The trapped flow of Fig.\ \ref{fig2}(a), which has a very small but finite value of $\Omega(0)$, does not have a fixed point anymore in a strict sense, but it stills resembles very closely the homogeneous flow. In this way, it is representative of a quasi-phase transition, that is expected for systems close to the thermodynamic limit. In Fig.\ \ref{fig2}(b), $\Omega(0)$ is larger by a factor of $10^3$, and the flow now differs significantly from the homogeneous one, although some characteristic features are still present. One should also note that
$M(0)$ now has to be changed over a much wider range than in (a) in order to produce the characteristic change in the large-$l$ behavior of the flow trajectories, i.e., switching from negative to positive final values $M(l^*)$.
This is in accordance with the expectation that the phase transition becomes more and more smeared out, as one moves away from the thermodynamic limit.

We now turn to the calculation of the grand thermodynamic potential that allows to calculate all thermodynamic properties of the trapped gas.
 In analogy with the symmetric phase of the homogeneous gas, within the
 quadratic approximation
\cite{WiegelJalickee}, the grand thermodynamic potential
in the symmetric phase of the trapped gas is given by
\begin{equation}
w = \frac{1}{\beta} \sum_{n_{1}, ... , n_{D}} \ln{\left[ 1-e^{-\beta
(E_{n} + E_{0}- \mu) } \right] }
= \frac{1}{\beta} \sum_{n=0}^{n_{\Lambda}}
 \sum^{'}_{ n_{1} ... n_{D} }
\ln{\left[ 1-e^{-\beta (E_{n}+E_{0}-\mu) } \right]}.
\end{equation}
In $D=3$, the constrained sum in the above equation can be determined easily
\cite{Grossmann1, Grossmann2, KetterlevanDruten} and $w$ reduces to
\begin{equation}
w = \frac{1}{\beta} \sum_{n=0}^{n_{\Lambda}}
 \left[ \frac{n^2}{2} + \frac{3n}{2} + 1 \right]
\ln{\left[ 1-e^{-\beta (E_{n}+E_{0}-\mu) } \right] }.
\label{z}
\end{equation}
Replacing the sum over $n$ with an integral in (\ref{z}),
shifting the origin so that the zero-point energy is eliminated from the
integrand, and using the generalized momentum $p = \hbar \Lambda e^{-l}$
instead of $n$ (compare with Sec.\ III A), we obtain in the thermodynamic limit
\begin{equation}
w =   \frac{1}{\beta} \int_{l' = 0}^{\infty} dl'
\frac 1 {8 \Omega(l')^3}
\ln{\left[ 1-e^{-b(l') (E_{>}-M(l')) } \right] }.
\label{zl}
\end{equation}
It is convenient to define an `intensive'
dimensionless thermodynamic potential $W(l) = \Omega(0)^3 \beta_{\Lambda} w(l)$ for which the scaling of the extensive potential $w$ in the thermodynamic limit is appropriately taken into account. Differentiating with respect to the continuous
variable $l$
we finally obtain the flow equation
\begin{equation}
\label{w}
\frac{dW(l)}{dl} =  \frac{e^{-6l}}{8b(0)}
\ln{[1 - e^{- b(l) [E_{>} - M(l)] } ] }.
\end{equation}
We note that in the above RG equation the chemical potential depends on $l$ and
therefore, through (\ref{trgeq}), on the
interaction. In other words, the
interaction enters the RG equation for the grand thermodynamic potential
implicitly through the non-trivial scaling of the chemical potential,
which is the main difference to the case of a trapped ideal gas.
In order to determine the grand thermodynamic potential,
we have to solve Eqs.\ (\ref{trgeq}) and (\ref{w})
with initial conditions $W(0)=0$, $M(0)\ll 1$,
$G(0) = \tilde G(0)b(0) \gg 1 $,
$b(0) \gg 1$, and $\Omega(0) \ll 1$.

The renormalization group equations (\ref{trgeq}) can be refined in a simple way
by means of a first order
$\varepsilon$-expansion \cite{Wilson,FisherStellenbosch}.
This method is expected to yield better
quantitative results in particular as far as critical properties are concerned.
For this purpose we set
$4-D=\varepsilon$ in (\ref{trgeq}) and formally treat this quantity as a small parameter.
We then expand systematically all quantities on the right hand sides of the RG equations (\ref{trgeq})
around $(M,\tilde{G}) = (0,0)$
up to second order in this small
parameter. Thereby one assumes that the quantities $M$ and $\tilde{G}$ remain of the order of
$\varepsilon$. Thus, we expand
the first equation of (\ref{trgeq}) up to first order in $M$ and the
second equation of (\ref{trgeq}) up to zeroth order in $M$.
As a consequence, in the thermodynamic limit we find
the modified RG equations
\begin{eqnarray}
\frac{d M(l)}{dl}
 & = &  2 M(l) - \tilde{ G}(l)\frac{\Omega_D}{(2\pi)^D} b(l)  \left\{ 2  N_m[b(l) E_{>}] + 2
 b(l) N_m[b(l) E_{>}] (1+N_m[ b(l) E_{>} ]) M(l) \right\},
\nonumber \\
\frac{d\tilde{G}(l)}{dl}  & = & \varepsilon \tilde{G}(l)  -
\tilde{G}(l)^2 \frac{\Omega_D}{(2\pi)^D}  b(l)\times\nonumber\\
& & \left\{ \frac{1+2N_m[b(l) E_{>} ]}{2E_{>}} + 4 b(l)
N_m[b(l) E_{>}] (1+N_m[b(l) E_{>}]) \right\}
\label{epsrgeq}
\end{eqnarray}
with $\Omega_D$ denoting the surface of a $D$-dimensional unit sphere (compare with Appendix A)
and with the modified Bose-distribution
$N_m(x) \equiv N_{BE}(x; \mu=0) = [\exp(x) - 1]^{-1}$. In the framework of the $\varepsilon$-expansion one sets
$D=4$ on the right hand side of Eqs.\ (\ref{epsrgeq}).
In three spatial dimensions these RG equations have to be solved for $\varepsilon =1$
together with the flow equation (\ref{w})
of the grand thermodynamic potential.

\section{Critical temperature}

Solving Eqs.\ (\ref{trgeq}) or (\ref{epsrgeq}) together with Eq.\
(\ref{w}) allows us to determine any thermodynamic property of an interacting,
trapped Bose gas at and above the BEC phase transition. First of all, let us summarize the results for the critical exponents which have been derived elsewhere \cite{StoofBijlsmaRen,FisherStellenbosch,Zinn} in connection with the study of the homogeneous flow. In the thermodynamic limit, the unstable fixed point of the flow equations (\ref{trgeq}) with $D=3$ is given by $(M^{*}, \tilde{G}^{*}) = (1/12, 5\pi^2/72)$. The critical exponent for the correlation length is $\nu=0.532$. Using first order $\varepsilon$-expansion, this fixed point is shifted to the position
$(M^*,\tilde{G}^*) = (\varepsilon/10, \varepsilon\pi^2/10)$ and its eigenvalues are given by
$\lambda_1 = 2 - 2\varepsilon/5$ and $\lambda_2 = - \varepsilon$ with $\varepsilon \equiv D-4 = 1$.
As a consequence the corresponding critical exponent for the correlation
length assumes the value $\nu = 1/\lambda_1 = 0.600$ which
is significantly closer to  the experimental value of
$\nu = 0.670$ \cite{Zinn}.

In the following, however, we wish to focus our attention on the dependence of the critical temperature on the interaction strength
in the thermodynamic limit.
The numerical investigation of critical properties requires us to find the critical flow trajectories in the
$(M, {\tilde G}, b)$ space, i.e., those trajectories that asymptotically reach the fixed point for $l\to\infty$. In Appendix B, we briefly comment on strategies for accomplishing this task. There exists a continuous family of different
critical trajectories, the so-called critical manifold, and each trajectory can be assigned a specific value for $a{\tilde N}^{1/6}/L$, i.e., a suitably normalized
scattering length, as we see below. Scanning the different critical trajectories thus allows us to explore a wide range of scattering lengths. The thermodynamic potential $W$ at criticality is then determined by integrating Eq.\ (\ref{w}) along a critical trajectory. Thereby, the initial and final values $[M(0), {\tilde G}(0), b(0)]$ and $[M(l_f), {\tilde G}(l_f), b(l_f)]$ of the trajectory need to be pushed sufficiently outwards so that the calculation of $W$ is converged.
Typically, this requirement implies that one has to choose $M(0)\ll 1$ and ${\tilde G}(0)b(0), b(0) \gg 1$. In this way, it is ensured that the cutoff $\Lambda$ corresponds to the largest energy in the problem, and relevant physical quantities such as $a{\tilde N}^{1/6}/L$ and the critical temperature become cutoff-independent.

Applying this procedure, it turns out, however, that for the largest scattering lengths shown in the figures below, convergence of $W$ cannot be achieved even if the initial conditions are pushed very far outwards. The results thus become somewhat cutoff-dependent. To explain this behavior, we argue that in this regime our RG model effectively describes a gas of hard spheres. We then attribute the cutoff-dependence to the fact that for strong interactions, the critical temperature no longer depends only on the scattering length, but is also sensitive to finer details of the potential. To establish the connection to a hard-sphere gas, we note that the bare two-body scattering length pertaining to a set of given initial conditions is obtained from the flow equation for $\tilde{g}(l) = g(l) e^{(D-2)l}$
in the limit of zero temperature and
zero chemical potential. The quantity $\tilde{g}(l)$ characterizes the renormalized interaction strength, after the trivial scaling has been removed.
Identifying the $s$-wave scattering length $a$ with the renormalized interaction strength $\tilde{g}(\infty)$
according to $\tilde{g}(\infty) = 4\pi a\hbar^2/m$
we obtain the relation
\begin{equation} \label{aLambda}
a \Lambda = \frac{\tilde G(0)b(0)}{4\pi + \frac{2}{\pi}\tilde G(0) b(0)}
\label{scatteringlength}
\end{equation}
for $D=3$.
This latter relation for the zero-energy $s$-wave scattering length
is in agreement with the Lippmann-Schwinger analysis \cite{Lippmann} for the $T$-matrix \cite{StoofBijlsmaRen}. We then compare this result to zero-energy two-body scattering with an interaction potential $V({\bf x}-{\bf x}') =  \Theta(R-|{\bf x}-{\bf x}'|)\hbar^2\kappa_0^2/2m_r$, i.e., a finite-height repulsive potential of radius $R$ \cite{Flu90}. Thereby $m_r=m/2$ denotes the reduced mass and $\kappa_0$ is a measure of the potential strength. If $\kappa_0R \gg 1$, the corresponding scattering length $a$ is given by $a/R = \kappa_0R/(1+\kappa_0R)$. With the help of the identifications $R = \pi/2\Lambda$ and $\kappa_0 R=\tilde G(0)b(0)/2\pi^2$, this expression can be mapped onto Eq.\ (\ref{aLambda}). In this way, we can relate the interaction used in (\ref{action}) to scattering between quasi-hard spheres, and $\pi/2\Lambda$ is interpreted as the sphere radius.

As emphasized in Sec.\ III B, in the thermodynamic limit the flow equations (\ref{trgeq}) and (\ref{epsrgeq}) reduce to the equations of an interacting homogeneous gas. Thus, all effects originating from the presence of the isotropic harmonic trap are contained in the flow equation (\ref{w}) for the dimensionless `intensive' thermodynamic potential $W(l)$. As $W(l)$ only depends on $M(l)$, but not on ${\tilde G}(l)$, the particle interactions affect the thermodynamic properties only through the non-trivial scaling of the dimensionless chemical potential $M(l)$. In the special case of negligible interaction, i.e. ${\tilde G}(l) = 0$, $M(l)$ scales trivially, i.e., $M(l) = M(0) e^{2l}$, so that Eqs. (\ref{trgeq}) and (\ref{w}) reproduce all thermodynamic properties of a non-interacting Bose gas in an isotropic harmonic trap within the framework of a grand canonical ensemble.

For a given critical trajectory with initial conditions $[M(0), {\tilde G}(0), b(0)]$, the calculation of the corresponding critical temperature $T_c$ proceeds as follows. With the critical trajectory, we associate the scaled particle number $s=-\partial W/\partial M|_{b(0), {\tilde G}(0)} = \Omega^{3} {\tilde N}$. A way to accurately calculate this quantity numerically is outlined in Appendix B. An ideal gas with the same value of $s = \Omega^{3} {\tilde N}$ has a critical temperature $T_c^0$ that is determined by $\beta_{\Lambda}k_B T_c^0 = [s\zeta(3)]^{1/3}$ with $\zeta(x)$ the Riemann zeta function \cite{KetterlevanDruten}. The ratio $T_c/T_c^0$ can thus be expressed as
\begin{equation}
T_{c}/T_{c}^{0} = \zeta(3)^{1/3}/(s^{1/3} b).
\label{ct}
\end{equation}
In our subsequent discussion we want to investigate the dependence of the critical temperature on the $s$-wave scattering length of the interacting Bose gas.
From Eq.\ (\ref{aLambda}) we find that a particularly convenient measure for the scattering length is the dimensionless parameter
\begin{equation}
a {\tilde N}^{1/6}/ L = s^{1/6} a \Lambda = s^{1/6}\frac{\tilde G(0)b(0)}{4\pi + \frac{2}{\pi}\tilde G(0) b(0)}.
\label{an}
\end{equation}
For each critical trajectory, we calculate $T_c/T_c^0$ and $a {\tilde N}^{1/6}/ L$, and in this way we obtain the dependence of the critical temperature on the scattering length.

\begin{figure}
\begin{center}
\includegraphics[width=8.6cm]{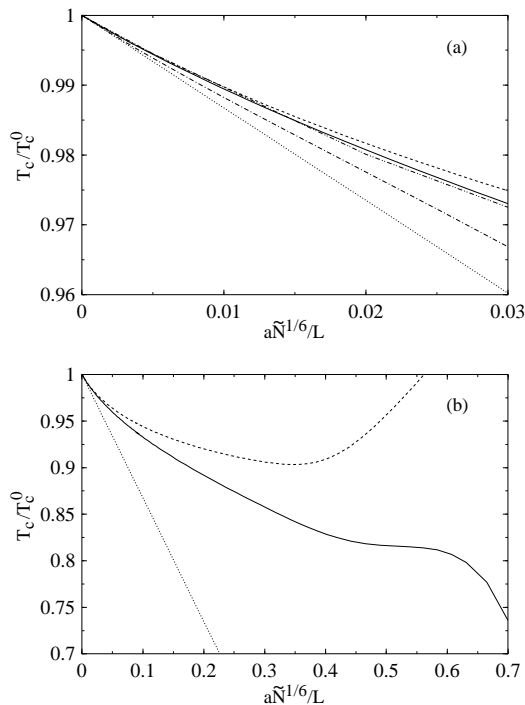}
\caption{Critical temperature as a function of the scattering length for the harmonically trapped interacting Bose gas. Bold curve: RG result after $\varepsilon$-expansion according to Eqs.\ (\protect\ref{epsrgeq}),
dashed: RG result without $\varepsilon$-expansion according to Eqs.\ (\protect\ref{trgeq}), dotted: mean field approximation \cite{GPS}.
The dot-dashed and dot-dot-dashed curves in (a) show the predictions of Refs.\ \cite{Arnold} and \cite{Houbiers}, respectively.
\label{fig3} }
\end{center}
\end{figure}

Figure \ref{fig3} summarizes our main numerical results for the critical temperature. Figure \ref{fig3}(a) shows the RG calculations according to Eqs.\ (\ref{epsrgeq}) (bold curve) and (\ref{trgeq}) (dashed) together with the mean field (MF) approximation \cite{GPS} (dotted) and the predictions of Refs.\ \cite{Arnold} (dot-dashed) and \cite{Houbiers} (dot-dot-dashed). It is remarkable that all results are within reasonable quantitative agreement. In particular, both RG results, which match rather closely for the values of $a {\tilde N}^{1/6}/ L$ shown in Fig.\ \ref{fig3}(a), lie above the MF approximation. This is generally expected \cite{Houbiers} as the critical fluctuations, that are neglected in the MF theory, should lead to an increase in $T_c$. In the limit of $a {\tilde N}^{1/6}/ L \to 0$, the RG curves tend to approach the MF result, and the ratio $(T_c^{RG}-T_c^{MF})/(T_c^0-T_c^{MF})$ seems to assume a constant value of about $7\%$. The prediction of \cite{Houbiers} closely agrees with the RG curves, whereas the result of \cite{Arnold} is intermediate between MF and RG.

In Fig.\ \ref{fig3}(b), we have extended our RG calculations to larger values of $a {\tilde N}^{1/6}/ L$. As mentioned above, for
$a {\tilde N}^{1/6}/ L$ larger than, approximately, 0.3, the quantitive results become somewhat cutoff-dependent. The general behavior of the curves, however, does not change. We now observe a profound difference between the treatments without and with $\varepsilon$-expansion. The former predicts an increase in $T_c/T_c^0$ for larger scattering lengths, whereas the latter indicates a decrease. On physical reasons, the second behavior appears to be more plausible, as for large particle interaction, an inhibition of Bose-Einstein condensation is expected.
On these grounds, the $\varepsilon$ expansion seems to be more reliable, but more work is certainly required to support these results further.

\begin{figure}
\begin{center}
\includegraphics[width=8.6cm]{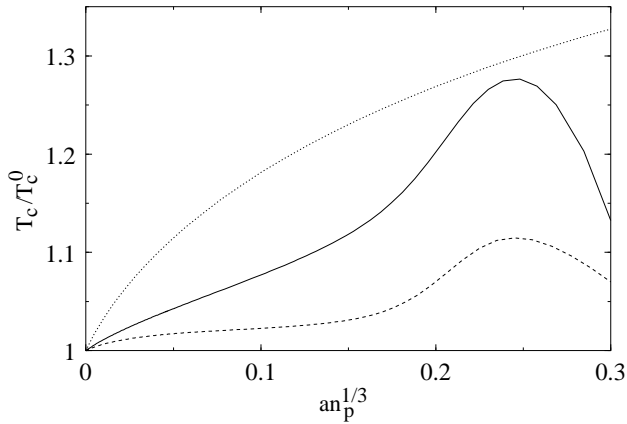}
\caption{Critical temperature as a function of the scattering length for the {\it homogeneous} interacting Bose gas. Bold curve: RG result after $\varepsilon$-expansion according to Eqs.\ (\protect\ref{epsrgeq}),
dashed: RG result without $\varepsilon$-expansion according to Eqs.\ (\protect\ref{trgeq}), dotted: RG result for symmetry-broken phase \cite{Alber}.
\label{fig4} }
\end{center}
\end{figure}

For the sake of comparison, we also show, in Fig.\ \ref{fig4}, the results for the RG treatment of the {\it homogeneous} interacting Bose gas. In this case, the relevant thermodynamic potential $W_{hom}$ is obtained from integrating \cite{StoofBijlsmaRen, Alber}
\begin{equation}
\frac{dW_{hom}}{dl} = \frac{1}{2\pi^2 b(0)}e^{-3l}
\ln{[1 - e^{- b(l) [E_{>} - M(l)] } ] },
\end{equation}
whereas the flow equations for $M(l)$ and ${\tilde G}(l)$ are still given by the thermodynamic limit of Eqs.\ (\ref{epsrgeq}) and Eqs.\ (\ref{trgeq}), respectively. In Fig.\ \ref{fig4}, the scattering length is scaled to the cubic root of the particle density $n_p$, so that $an_p^{1/3}=s_{hom}^{1/3}{\tilde G(0)}b(0)
/(4\pi+2\tilde G(0)b(0)/\pi)$ with $s_{hom}=-\partial{W_{hom}}/\partial M|_{b(0), {\tilde G}(0)}$.
The scaled critical temperature is obtained from $T_c/T_c^0=
[\zeta(3/2)/s_{hom}]^{2/3}/2\pi b(0)$. Together with the results for the symmetric phase we also display the calculation for the symmetry-broken phase as described in Ref.\ \cite{Alber}. For small $an_p^{1/3}$ it was established in \cite{Alber} that the computation in the symmetry-broken phase predicts a linear dependence of the critical temperature on $a$, i.e., $T_c/T_c^0= C an_p^{1/3}$ with $C\approx 3.4$. For the symmetric phase without and with $\varepsilon$-expansion, we also find an approximately linear behavior with constants $C=1.0$ and 1.4, respectively. All these results fit reasonably well into the `zoo' of current predictions \cite{Andersen03}. In fact, the value for $C$ obtained from the symmetric phase with first-order $\varepsilon$-expansion matches rather closely the results of Refs.\ \cite{KasProSvi01,ArnMoo01} which may be regarded as the best current estimates \cite{Shl03}.

It is also of interest to consider the regime of large scattering lengths. In the symmetry-broken phase, the calculation predicts a monotonous increase of $T_c$ with the scattering length, a behavior which is probably unphysical. Remarkably, however, both curves for the symmetric phase show a maximum and a subsequent decrease in the critical temperature. The location and in particular the height of the maxima are somewhat cutoff-dependent, but altogether they do not differ too strongly from the results displayed in Ref.\ \cite{Cep97}. As outlined above, we attribute the cutoff-dependence to the fact that in this regime our RG model effectively describes a gas of quasi-hard spheres. For strong enough interactions the critical temperature cannot be parametrized by the scattering length alone, but also depends on the finer details of the potential. Varying the cutoff, e.g., amounts to changing the radius of the hard spheres.

\section{Summary and conclusions}

We have applied the Wilsonian momentum-shell renormalization group technique to the harmonically trapped interacting Bose gas. By integrating out small energy shells in the partition function and applying the $\varepsilon$-expansion, we have obtained flow equations for the thermodynamic variables. In the thermodynamic limit, the flow equations for the chemical potential and the interaction potential reduce to the corresponding relations for the homogeneous interacting Bose gas. The presence of the trap becomes manifest only through the modified flow equation for the grand thermodynamic potential. From the flow equations, we have calculated the transition temperature as a function of the scattering and found our results in good agreement with previous approaches.
In future work, we plan to extend the methods presented here to more general potential shapes in order to assess the general applicablity of our approach.

\begin{acknowledgments}
This work is supported by the Deutsche Forschungsgemeinschaft through the Forschergruppe ``Quantengase." Stimulating discussions with G.\ Shlyapnikov are gratefully acknowledged. G.\ M.\ wishes to thank Prof.\ I.\ J.\ R.\ Aitchinson for many valuable discussions.
\end{acknowledgments}

\begin{figure}
\begin{center}
\includegraphics[width=8.6cm]{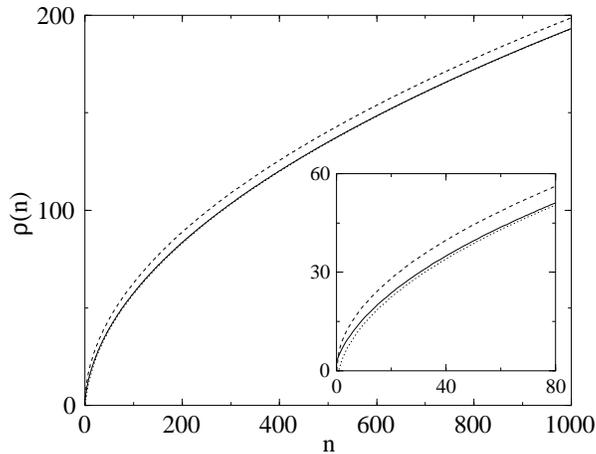}
\caption{Numerical calculation of $\rho(n)$ of Eq.\ (\ref{rho}) (full curve) and
approximations $2\pi \sqrt{n}$ (dashed) and $2 \pi \sqrt{n} - 5.701$ (dotted).
\label{fig5}}
\end{center}
\end{figure}

\appendix
\section*{Appendix A}

In this Appendix, we discuss a method for calculating the constrained sum
$\rho(n)$ of Eq.\ (\ref{rho}) in the limit
of large quantum numbers and wide traps, i.e., $n \gg 1$ and
$\beta \hbar \omega \to 0$. In this case, the difference between the energy levels of the harmonic oscillator vanishes and we can therefore replace the sums in
(\ref{rho}) by integrals.
Furthermore, in the limit of large values of $n$ the dominant contribution to Eq.\ (\ref{rho})
comes from values $n_1, ...,n_D \gg 1$ so that we obtain the approximate relation
\begin{eqnarray}
\int dn~\rho(n) &=&
\int dn~\int^{'} dn_{1} ... dn_{D}~ \frac{1}{ \sqrt{n_{1}
... n_{D} } }= 2^{D} \int d\sqrt{n_{1}} ... d\sqrt{n_{D}}.
\nonumber
\end{eqnarray}
Here
we have converted the constrained integral over
$n_{j}, j=1,...,D$ and the integration over $n$ into an unconstrained
integration over $n_{j}$ and then changed variables to
$\sqrt{n_{j}}$.
Next we define the vector $ {\bf r}= (\sqrt{n_{1}}, ... , \sqrt{n_{D}})$
which fullfils $r^2=n_{1}+...+n_{D}=n$. This implies that 
\begin{eqnarray}
2^{D} \int d\sqrt{n_{1}} ...  d\sqrt{n_{D}}= 2^{D} \frac{1}{2^{D}} 
\int~\Omega_{D}~r^{D-1}~dr &=& \int~\frac{\Omega_{D}}{2}~n^{\frac{D}{2}-1}~dn
\nonumber
\end{eqnarray}
where $\Omega_{D}=2 \pi^{D/2} / \Gamma(D/2)$ is the surface of a
$D$-dimensional unit sphere, and $\Gamma(x)$ denotes the Gamma function. The factor $1/2^{D}$ is due to the fact that the
integration variables take only positive values.
We have thus derived the main result
\begin{eqnarray}
\rho(n) &=& \frac{\Omega_{D}~ n^{\frac{D}{2}-1}}{2}
\nonumber
\end{eqnarray}
which is valid in the limit $n\to \infty$.
For $D=3$ we have
$\Omega_{3}=4 \pi$ and $\rho(n)=2 \pi \sqrt{n}$, 
so that in the limit $\Omega(l) = 1/(L\Lambda e^{-l})^2 \gg 1$ the quantity $d(l)$ of Eq.(\ref{density}) 
reduces to the result $d_{\rm h}=1/2
\pi^2$ and the RG equations (\ref{trgeq}) reduce to the ones for a homogeneous interacting Bose
gas in the absence of a trap. Figure \ref{fig5} shows a diagram of $\rho(n)$ together with the approximation $2\pi\sqrt{n}$. It is also possible to compute
 numerically the next-order correction in the high-$n$ expansion,
$2 \pi \sqrt{n} - 5.701$, which is almost indiscernible from the exact result, except near $n=0$ (see inset).

\section*{Appendix B}

In this Appendix, we briefly discuss some of our numerical methods. We first describe how we determine the
derivative $\partial W/\partial{M}\mid_{b(0),\tilde G(0)}$ which is required to calculate the quantities
$T_c/T_c^0$ and $a{\tilde N}^{1/6}/L$ of Eqs.\ (\ref{ct}) and (\ref{an}),
respectively. From these quantities, we obtain Figs.\ \ref{fig3} and \ref{fig4} showing the
dependence of the critical temperature on the scattering length.

A straightforward approach would consist in using finite differences, i.e.,
$\partial W/\partial{M}\mid_{b(0),\tilde G(0)} \approx [ W(M_0+\delta M) -  W(M_0)]/\delta M$
with $M_0$ the chemical potential for which the derivative is required
and $\delta M$ a small variation. However, this method turns out to not
produce well-defined results, in particular for small values of
$a{\tilde N}^{1/6}/L$. Therefore, we apply a scheme which is closely related
to the method of linear stability analysis frequently used in the theory of
dynamical systems \cite{Gut90}. In this scheme, one propagates the flow trajectory,
from which $W(M_0)$ is calculated, along with an infinitesimal linear
deviation from the initial conditions.

To be specific, we consider the evolution of quantities $\delta_Y X(l)$
with $X,Y=M$ or $\tilde{G}$. These quantities are to be interpreted as
$\delta_Y X(l) = \delta X(l)/\delta Y(0)$, i.e., they describe the change
$\delta X(l)$ of a flow trajectory if the initial condition is changed 
infinitesimally 
by $\delta Y(0)$. If we  write the flow equations (\ref{trgeq}) or (\ref{epsrgeq}) symbolically as
\begin{eqnarray*}
\frac{dM}{dl} & = & f_M(M, \tilde{G}), \\
\frac{\tilde{G}}{dl} & = & f_{\tilde{G}}(M, \tilde{G}),
\end{eqnarray*}
then the deviations $\delta_Y X(l)$ obey the differential equations
\begin{displaymath}
\frac d {dl}\left(\begin{array}{c}
\delta_M M \\[0.1cm] \delta_M \tilde{G} \end{array}\right) =
\left(\begin{array}{cc}
\frac{\partial f_M}{\partial M} & \frac{\partial f_M}{\partial \tilde{G}} \\[0.1cm]
\frac{\partial f_{\tilde{G}}}{\partial M} & \frac{\partial f_{\tilde{G}}}{\partial \tilde{G}}
\end{array}\right)
\left(\begin{array}{c}
\delta_M M \\[0.1cm] \delta_M \tilde{G} \end{array}\right)
\end{displaymath}
with the initial conditions $\delta_M M(0) = 1$ and $\delta_M \tilde{G}(0) = 0$.
The matrix of partial derivatives has to be evaluated along the
original flow trajectory. Writing Eq.\ (\ref{w}) symbolically as
$dW(l)/dl = F(M)$, the required derivative
$\partial W/\partial{M}\mid_{b(0),\tilde G(0)}= \partial W/\partial{M}\mid_{b(0),\tilde G(0)}(l\to\infty)$ is
finally obtained from
\begin{displaymath}
\frac d{dl}\left(\frac{\partial W}{\partial{M}}\mid_{b(0),\tilde G(0)}\right) =
\frac{\partial F}{\partial M}\delta_M M
\end{displaymath}
with the initial condition $\partial W/\partial{M}\mid_{b(0),\tilde G(0)}=0$.
We have verified that the results from this method agree with the
finite differencing scheme in regimes where the latter is applicable, e.g.,
for larger values of $a{\tilde N}^{1/6}/L$.

We now outline how to find the critical trajectories, i.e., those trajectories that asymptotically reach the fixed point. A systematic and reliable way to determine the critical $M(0)$ for given $[\tilde G(0),b(0)]$ consists in propagating trajectories with increasing $M(0)$, starting from $M(0)=0$. For these trajectories $M(l\to\infty)\to -\infty$ initially. However, for a large enough $M(0)=M_1$, one eventually finds $M(l\to\infty)\to 1/2$. The crossover between these two opposite behaviors occurs at the critical trajectory. The critical $M(0)$ is thus bracketed by 0 and $M_1$ and can now be found, e.g., via bisection. Another method, which is computationally more efficient but works well only for smaller scattering lengths, is based on backwards propagation from the non-trivial fixed point. To this end, one first finds the critical manifold of the
fixed point in the linearization approximation.
For a given $b(l_f) \ll 1$ and
starting from points $(M(l),\tilde{G}(l))$ on this critical manifold, very near
the fixed point, we propagate the thermodynamic
limit of Eqs.\ (\ref{trgeq}) or (\ref{epsrgeq})
backwards up to a value $l=l_{i}$ for which $\tilde G(l_{i})$ is sufficiently large so that it
is ensured that $a\Lambda < 1$. In this way we find the initial conditions
of a critical trajectory $[M(l_{i}), \tilde{G}(l_{i}), b(l_{i})]$.
Further critical trajectories are computed by repeating the above procedure
using  different values of $b(l_f)$.


\begin{thebibliography}{10}

\bibitem{AndEnsMat95} M.~H.\ Anderson, J.~R.\ Ensher, M.~R.\ Matthews, C.\ Wieman, and E.~A.\ Cornell, Science {\bf 269}, 198 (195).

\bibitem{DavMewAnd95} K.~B.\ Davis, M.~O.\ Mewes, M.~R.\ Andrews, N.~J.\ van Druten, D.~S.\ Durfee, D.~M.\ Kurn, and W.\ Ketterle, Phys.\ Rev.\ Lett.\ {\bf 75}, 3969 (1995).

\bibitem{BraSacTol95} C.~C.\ Bradley, C.~A.\ Sackett, J.~J.\ Tollett, and R.~G.\ Hulet, Phys.\ Rev.\ Lett.\ {\bf 75}, 1687 (1995).

\bibitem{Kadanoff}
L.~P.\ Kadanoff, {\em Statistical Physics: Statics, Dynamics and
  Renormalization} (World Scientific, Singapore, 2001).

\bibitem{Huang}
K.\ Huang, {\em Statistical Mechanics} (Wiley, New York, 1987).


\bibitem{StoofBijlsmaRen}
M. Bijlsma and H.~T.~C.\ Stoof, Phys.\ Rev.\ A {\bf 54}, 5085 (1996).

\bibitem{BraatenNieto}
E.\ Braaten and A.\ Nieto, Phys.\ Rev.\ B {\bf 55}, 8090 (1997).

\bibitem{Andersen}
J.~O.\ Andersen and M.\ Strickland, Phys.\ Rev.\ A {\bf 60},  1442  (1999).

\bibitem{Alber}
G.\ Alber, Phys.\ Rev.\ A {\bf 63}, 023613 (2001).

\bibitem{Crisan}
M.\ Crisan, D.\ Bodea, I.\ Grosu, and I.\ Tifrea, J.\ Phys.\ A {\bf 35},  239  (2002).

\bibitem{AlberMetikas}
G.\ Alber and G.\ Metikas, Appl.\ Phys.\ B {\bf 73},  773  (2001).

\bibitem{Metikas}
G.\ Metikas and G.\ Alber, J.\ Phys.\ B {\bf 35},  4223  (2002).

\bibitem{CreswickWiegel}
R.~J.\ Creswick and F.~W.\ Wiegel, Phys.\ Rev.\ A {\bf 28},  1579  (1983).

\bibitem{FisherHohenberg}
D.~S.\ Fisher and P.~C.\ Hohenberg, Phys.\ Rev.\ B {\bf 37},  4936  (1988).

\bibitem{Weichman}
P.~B.\ Weichman, Phys.\ Rev.\ B {\bf 38},  8739  (1988).

\bibitem{Nelson}
D.~R.\ Nelson and H.~S.\ Seung, Phys.\ Rev.\ B {\bf 39},  9153  (1989).

\bibitem{Kolomeisky1}
E.~B.\ Kolomeisky and J.~P.\ Straley, Phys.\ Rev.\ B {\bf 46},  11749  (1992).

\bibitem{Kolomeisky2}
E.~B.\ Kolomeisky and J.~P.\ Straley, Phys.\ Rev.\ B {\bf 46},  13942  (1992).

\bibitem{Wilson} K.~G.~Wilson and J.~Kogut, Phys.\ Rep., Phys.\ Lett.\ {\bf 12}, 75 (1974). 

\bibitem{Feynman} R.~P.~Feynman, {\em Statistical Mechanics} (Benjamin, Reading, MA, 1972).

\bibitem{NotesStoof}
H.~T.~C. Stoof, {\em Field theory for trapped atomic gases}, Lecture Notes for
  Les Houches Summer School on Coherent Atom Waves, e-print cond-mat/9910441, (1999).

\bibitem{FetterWalecka}
A.~L.\ Fetter and J.~D.\ Walecka, {\em Quantum Theory of Many-Particle Systems}
  (McGraw-Hill, New York, 1971).

\bibitem{SchwingerFock}
J.\ Schwinger, Phys.\ Rev.\ {\bf 82},  664  (1951).

\bibitem{Aitchison}
I.~J.~R.\ Aitchison, Acta Phys.\ Pol.\ B {\bf 18},  191  (1987).

\bibitem{AML}
I.~J.~R.\ Aitchison, G.\ Metikas, and D.~J.\ Lee, Phys.\ Rev.\ B {\bf 62},  6638
  (2000).

\bibitem{BerryMount}
M.~V.\ Berry and K.~E.\ Mount, Rep.\ Prog.\ Phys.\ {\bf 35},  315  (1972).


\bibitem{Gleisberg}
F.\ Gleisberg, W.\ Wonneberger, U.\ Schl\"oder, and C.\ Zimmermann, Phys.\ Rev.\ A {\bf 62},  063602  (2000).

\bibitem{Grossmann1}
S.\ Grossmann and M.\ Holthaus, Z.\ Naturforsch.\ A {\bf 50},  921  (1995).

\bibitem{Grossmann2}
S.\ Grossmann and M.\ Holthaus, Phys.\ Lett.\ A {\bf 208},  188  (1995).

\bibitem{KetterlevanDruten}
W.\ Ketterle and N.~J.\ van Druten, Phys.\ Rev.\ A {\bf 54},  656  (1996).


\bibitem{WiegelJalickee}
F.~W.\ Wiegel and J.~B.\ Jalickee, Physica {\bf 57},  317  (1972).

\bibitem{FisherStellenbosch}
M.~E. Fisher, {\em Scaling, Universality and Renormalization Group Theory},
  Lectures presented at the ``Advanced Course on Critical Phenomena" held at The
  Merensky Institute of Physics, University of Stellenbosch, South Africa,
  (1982).

\bibitem{Zinn}
J. Zinn-Justin, {\em Quantum Field Theory and Critical Phenomena} (Oxford
  University Press, Oxford, 1989).

\bibitem{Lippmann}
B.~A.\ Lippmann and J.\ Schwinger, Phys.\ Rev.\ {\bf 79},  469  (1950).

\bibitem{Flu90} S.\ Fl\"ugge, {\it Rechenmethoden der Quantentheorie}, (Springer, Berlin, 1990).

\bibitem{GPS}
S.\ Giorgini, L.~P.\ Pitaevskii, and S.\ Stringari, Phys.\ Rev.\ A {\bf 54},  R4633 (1996).

\bibitem{Arnold}
P.\ Arnold and B.\ Tom\'asik, Phys.\ Rev.\ A {\bf 64},  053609  (2001).

\bibitem{Houbiers}
M.\ Houbiers, H.~T.~C.\ Stoof, and E.~A.\ Cornell, Phys.\ Rev.\ A {\bf 56},  2041 (1997).

\bibitem{Andersen03} J.~O.\ Andersen, e-print cond-mat/0305138.

\bibitem{KasProSvi01} V.~A.\ Kashurnikov, N.~V.\ Prokof'ev, and B.~V.\ Svistunov, Phys.\ Rev.\ Lett.\ {\bf 87}, 120402 (2001).

\bibitem{ArnMoo01} P.\ Arnold and G.~D.\ Moore, Phys.\ Rev.\ E {\bf 64}, 066113 (2001).

\bibitem{Shl03} G.~V.\ Shlyapnikov, {\it Condensed matter approaches for quantum gases}, in Poincar\'e Seminar 2003 (Birkh\"auser, Basel, 2003).

\bibitem{Cep97}  P.\ Gr\"uter, D.\ Ceperley, and F.\ Lalo\"e, Phys.\ Rev.\ Lett.\ 79, 3549 (1997).

\bibitem{Gut90} M.~C.\ Gutzwiller, {\it Chaos in classical and quantum mechanics} (Springer, New York, 1990).






\end{thebibliography}
\end{document}